\documentclass[
a4paper
]{ceurart}

\usepackage{listings}
\usepackage[T1]{fontenc}
\usepackage{lmodern}

\begin{document}

\copyrightyear{2024}
\copyrightclause{Copyright for this paper by its authors.
  Use permitted under Creative Commons License Attribution 4.0
  International (CC BY 4.0).}

\conference{SeMatS 2024: International Workshop on Semantic Materials Science, September 17--19, 2024, Amsterdam}

\title{Enhancing Semantic Interoperability Across Materials Science With HIVE4MAT}

\author[1]{Jane Greenberg}[%
]
\address[1]{Metadata Research Center, College of Computing and Informatics, Drexel University,
  3675 Market St., Philadelphia, PA 19104, USA}

\author[1]{Kio Polson}[%
orcid=0009-0002-3087-0739
]

\author[1]{Scott McClellan}[%
url=http://conceptbase.sourceforge.net/mjf/,
]

\author[1]{Xintong Zhao}[%
]

\author[1]{Alex Kalinowski}[%
]

\author[1]{Yuan An}[%
]


\begin{abstract}
HIVE4MAT is a linked data interactive application for navigating ontologies of value to materials science. HIVE enables automatic indexing of textual resources with standardized terminology. This article presents the motivation underlying HIVE4MAT, explains the system architecture, reports on two evaluations, and discusses future plans.
\end{abstract}


\maketitle

\section{Introduction}

Data infrastructure advancements over the last two decades have supported the development of a rich ecosystem of machine readable semantic vocabularies across many domains. These advances have enabled a number of  materials science initiatives, e.g.\cite{bayerlein2024pmd,zhao2018nanomine} to leverage individual ontologies for data organization, access, interoperability, and support for the FAIR data principles\cite{wilkinson2016fair}. Even with this progress, limited attention has been directed to exploring the broad array of ontologies and other semantic systems both within materials science and across interconnected disciplines. One likely reason for this condition is the general evolution of ontologies. Ontological development and use-case scenarios represent a first-phase priority, while human and financial resources for exploring the larger spectrum of semantic systems require more time and investment. Another factor is limited availability of applications that provide researchers and data custodians capacity to easily explore and use of the broader spectrum of relevant semantics systems. Research is necessary to allow for greater exploration of a wide array of semantic systems of value to materials science researchers for organizing and providing access to their data assets and other scientific and scholarly outputs. 

The Helping Interdisciplinary Vocabulary Engineering for Materials Science (HIVE4MAT) application seeks to address this challenge. HIVE4MAT allows exploration multiple ontologies in a single environment through search, automatic indexing, as well as targeted browsing of an individual ontology's simple, taxonomic structure. This paper reports on HIVE4MAT research development, the support for breaking down research silos of terminologies, baseline evaluations, and future plans.

\section{Case for HIVE4MAT}

Technical advanecments have greatly accelerated the capacity for machine readable, shared semantic systems, including ontologies. This progress has largely been motivated by Berners-Lee’s vision of the Semantic Web as a world of linked data\cite{berners2001publishing} together with the development and adoption of encoding standards underlying machine-readable ontological systems\cite{bizer2009emerging}. Primary standards  supported by the World Wide Web Consortium (W3C) include the the Resource Description Framework (RDF) data model and the Web Ontology Language (OWL). Another standard is the Simple Knowledge Organization Systems (SKOS) for representing terminological and classificatory Knowledge Organization Systems (KOS), such as thesauri, taxonomies, classification schemes, controlled vocabularies, and keyword lists. SKOS is defined using an RDF schema and supports linked data and the Semantic Web ecosystem\cite{Miles:09:SSK}. While less logically complex than OWL, the SKOS data model can better support collective analysis of multiple ontologies through a set of taxonomic relationships among concepts.

Over the past several decades, the availability of these encoding standards, examples of ontology development in biology and bio-medicine, and, more recently, national and international data-driven initiatives have motivated ontology development in materials science. For example the U.S. Materials Genome Initiative\cite{DimaAlden2016IIft,alma991010609049704721} calls for accelerated, data-driven innovation and discovery of materials which are less environmentally harmful. Robust, standardized metadata and the use of ontologies are key to this aim. Another example is the extensive effort of the European Materials Modelling Council\cite{GoldbeckG2019Arla} and the support for the  Elementary Multiperspective Material Ontology (EMMO). Other developments include the global adoption of the FAIR principles, advancements in knowledge graph research\cite{VoigtSvenP.2021Mgo,DavidMrdjenovich2020pAKG}, and the rapid growth of artificial intelligence\cite{MaBoran2023MUoP}.

Advancement in the overall ontology landscape has also involved development of trusted registries for sharing these semantic systems for broader community use. The biomedical and biology communities have led in this area with the National Center for Biological Ontologies (NCBO) Bioportal\footnote{\url{http://bioportal.bioontology.org/}}\cite {whetzel2011bioportal} and the OBO foundry\footnote{\url{http://www.obofoundry.org/}}\cite{smith2007obo}. These registries have allowed biomedical, bio-science, earth science, and other discipline researchers a greater opportunity to explore and use multiple ontologies. In fact, a number of ontologies found in these registries are applicable to materials science research, given its interdisciplinary nature. There has also been increased, complementary activity to deploy materials science registries, with NOMAD\cite{draxl2018nomad}, NIST Materials Registry\footnote{\url{https://materials.registry.nist.gov/}}\cite{PlanteRaymondL2021IaRF}, the Industrial Ontology Foundry\footnote{\url{https://www.industrialontologies.org/}}, and most recently MatPortal\footnote{\url{https://matportal.org/ontologies}}, based on The Bioportal’s technology\cite{jonquet2023ontoportal}.

Even with these developments, materials science ontology advancements have primarily been siloed, with use of an individual ontology to support and organize the underlying a specific data model of a single project. Examples include Nanomine\cite{zhao2018nanomine}, NOMAD vocabulary\cite{DraxlClaudia2019TNlf}, and Process Material Digital Core ontology\cite{bayerlein2024pmd}. These developments have been significant for advancing the semantic space in materials science. However, leveraging a broader collection of ontologies in a single workflow will enhance semantic interoperability and support of the FAIR principles both within materials science and in relation to other disciplines. To this end, there is a need for applications that allow researchers and data custodians to explore and implement multiple ontologies and determine their applicability. This is a key goal of the HIVE4MAT application presented in this paper. 

\section{Developing Solution: HIVE4MAT}
\subsection{HIVE4MAT Overview}
HIVE4MAT is a linked data, interactive application for navigating ontologies, automatically indexing text, and generating metadata with standardized terms drawn from ontologies. HIVE4MAT is led by the Metadata Research Center at Drexel University. This initiative is also connected with the National Science Foundations Harnessing the Data Revolution's Institute for Data Driven Dynamical Design (NSF/HDR-ID4)\footnote{\url{https://www.mines.edu/id4/}}, where one of the Institute's aims is to facilitate greater interoperability and communication among different facets of materials science. The intended user audience for HIVE4MAT includes material science researchers, curators, and data custodians; although, researchers who regularly interact with ontologies may also find this application of interest. HIVE4MAT  features simplify ontology use for semantic representation and curation activities, as the chief audience is generally not sophisticated ontology developers. 

The main functions of HIVE4MAT is to provide users with standardized semantics terminology for metadata representations  drawing from multiple ontologies and semantic systems. HIVE4MAT does not aim to support reasoning  and complex relationships represented in OWL, but instead retains basic taxonomic equivalency, hierarchy, and associative relationships\cite{national_information_standards_organization_guidelines_2010}. While HIVE4MAT does not perform sophisticated reasoning, the SKOS encoded relationships offer a pragmatic approach for navigating and comparing multiple ontologies in a single sequence, and a good place to initiate ontology exploration. Finally, while HIVE4MAT's use of SKOS collapses OWL ontologies, users of this application are able to find, explore, and use the full OWL ontology through other systems. 

\subsection{Functionality}

The HIVE4MAT application has three main user features: 1. Navigation, 2. Search, and 3. Indexing. These three features are discussed below in more detail.

\subsubsection{Navigation}

The navigation feature allows a user to select and explore an ontology tree hierarchically from the top level concepts down to any child or descendant concepts. Ontologies included in HIVE4MAT are converted from any RDF-based format into the SKOS schema. While this translation may eliminate some of the initial OWL functionality, it does not reduce users' ability to effectively explore ontological concepts within a simple taxonomic tree structure. Moreover, the use of a single standard reduces the ambiguity between super- and sub-class relationships\cite{baker2013key}. Users can launch their ontology navigation at an ontology's top level concept, a node concept, or a leaf concept. Once the user selects a concept, they may view the full SKOS encoded metadata for that particular concept. The HIVE4MAT SKOS includes the following attributes: 

\begin{itemize}
\item Preferred label (SKOS:preflabel)
\item URI
\item Alternate label (SKOS:altlabel)
\item Notes
\item SubClassOf (SKOS:broader)
\item SuperClassOf (SKOS:narrower)
\item Related Concepts (SKOS:related)
\end{itemize}

If the user identifies a concept that they plan to use in their metadata, HIVE4MAT allows the user to copy the concept and the preferred encoding. HIVE4MAT supported metadata encodings include:

\begin{itemize}
\item JSON-LD
\item SKOS RDF/XML
\item Dublin Core XML
\item ``Plain'' XML
\end{itemize}

These metadata encodings are components of HIVE4MAT's additional features  described below. 

\subsubsection{Searching}

HIVE4MAT allows a user to search for a concept within an ontology or across multiple ontologies. The search feature permits the user to select one or a set of ontologies to search, and then search to retrieve relevant terms in the selected ontology or ontologies. Following the search, HIVE4MAT returns a list of relevant terms grouped by ontology for the user to browse. A HIVE4MAT search query is run against the following concept fields:  preferred label (SKOS:preflabel) the alternate label (SKOS:altlabel), and the annotations (SKOS:scopenote). As a result, retrieved concepts may not contain the initial search term entered by the user as a preferred term. This helps users identify a more robust terminology for their metadata record.

As with the navigation feature, once the list of relevant terms is displayed, the user can click on any of the terms and see the associated metadata of the concept. The user can also copy the metadata for one concept at a time in the encoding formats listed earlier.

\subsubsection{Automatic Indexing}

HIVE4MAT also has an automatic indexing feature for processing text and assisting users in the selection of standard ontology terms for area-specific descriptive metadata. The HIVE4MAT uses Natural Language Processing (NLP) and then maps keyword phrases to the the selected ontologies. A user can select to use either the Rapid Automatic Keyword Extraction (RAKE)\cite{rose2010automatic} or Yet Another Keyword Extractor (YAKE)\cite{campos2020yake} algorithm. Following the indexing sequence, the user (e.g., a researcher, curator, data custodian) is presented with indexing results and selects the desired terms from one or more ontologies for the metadata representation they are creating.

\begin{figure}[h!]
    \centering
    \includegraphics[height=9.0cm]{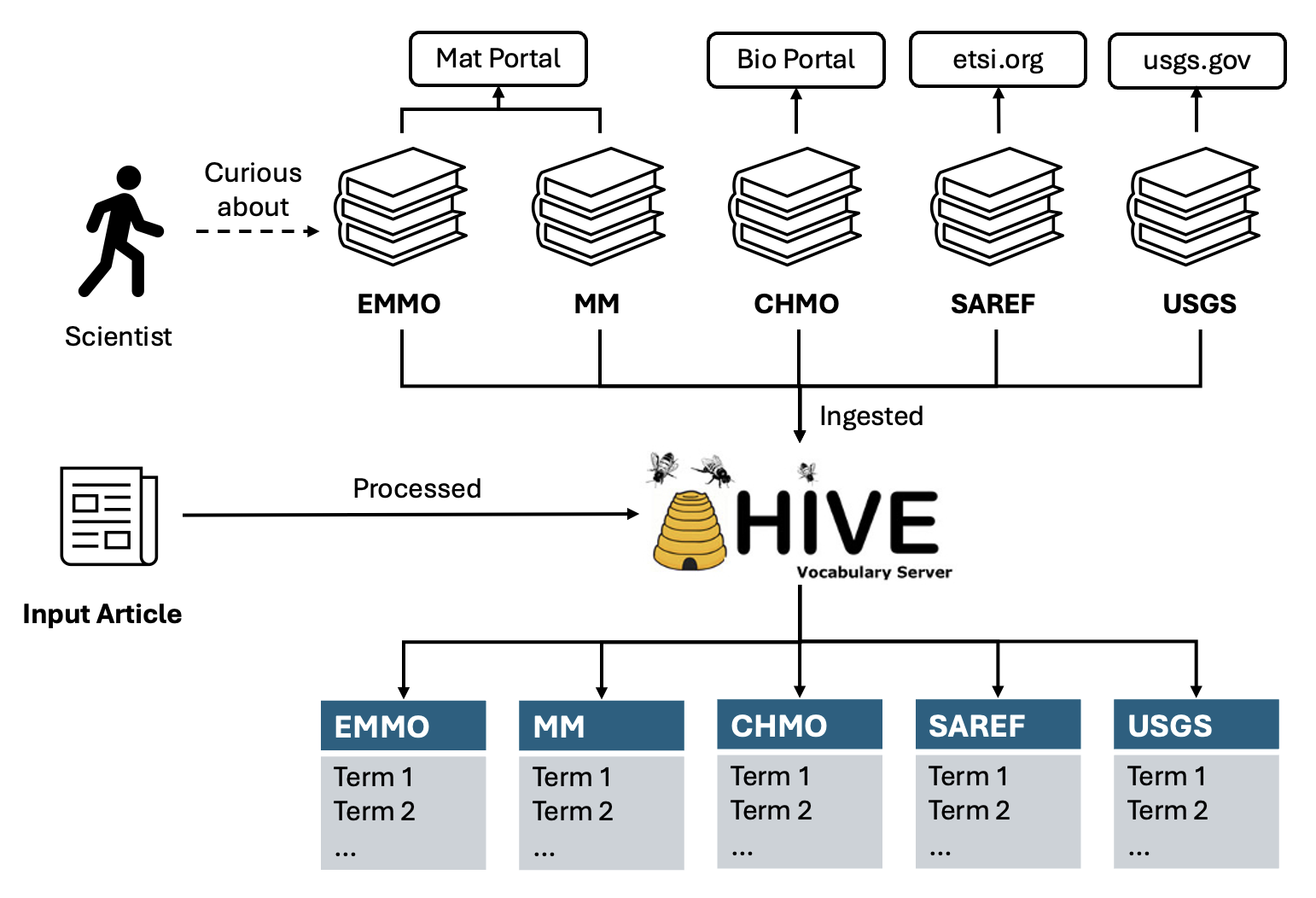}
    \caption{HIVE4MAT example indexing workflow}
    \label{fig:workflow}
\end{figure}

Figure \ref{fig:workflow} provides an overview of HIVE4MAT workflow, and a detailed scenario follows here: The user first selects which ontologies they would like to consider for the automatic indexing sequence. Next the user can either upload a text file, an MSWord document, or a PDF which HIVE4MAT converts to text can then index. Alternatively, the user may type in a URI of a web page or another accessible digital resource, and HIVE4MAT will then scrape for the textual content of the page for automatic indexing. HIVE4MAT provides options for the user to select and modify algorithm settings and word count lengths. The last step in the indexing workflow requires the user to click the indexing button. This step initiates automatic indexing of text that has been either uploaded or identified with a URI. HIVE4MAT also has a script to support batch up-loading and indexing if there is collection of documents to process. 

HIVE4MAT takes several seconds to index against multiple ontologies. This greatly expedites the process of having to index an article against one ontology at a time. We should note that the larger ontologies can slow down this process to anywhere between seven and thirty seconds, although this time lag is still more expedient than indexing a resource against each ontology, one at a time; and our development team continues to work on improvements in this area.

After the indexing activity, candidate terms are displayed and grouped together by ontology. Relevancy for candidate terms is indicated by font size and order. The user can further manipulate the results, as there are four different sort and display options. The automatic indexing results focus only on terms in in the SKOS preferred label, and the alternate labels and notes are not part of the sequence. Finally, as with the search and navigation features, users can click on any term and the associated metadata for that concept will be displayed, which can then be copied using the encodings listed above.
\subsubsection{Application Architecture}

HIVE4MAT is written in Python 3, and it follows the framework of general HIVE, and the ontologies are stored as SQL databases. With the exception of three ontologies generated in Protege, the selection of ontologies currently in HIVE4MAT were retrieved from the Bioportal \cite{whetzel2011bioportal} and MatPoral \footnote{\url{https://matportal.org/ontologies}}. They were converted to SKOS on loacal computers using one of the team's python scripts, and uploaded into HIVE4MAT one at a time. A key goal is to automate this process. 

\section{Evaluation and Current Development}

Over the last several years, the HIVE4MAT team has conducted several evaluations that impact current development works. Two key evaluations efforts pursued with materials science researchers are shared here.

The first evaluation was 2021 study comparing the performance of HIVE4MAT, as a basic knowledge extraction automatic indexing application to MatScholar, a named entity recognition (NER) application\cite{zhao2021exploratory}. A sample of 60 abstracts from inorganic materials research articles were processed through both HIVE4MAT, which at the time used classic RAKE (rapid automatic keyword extraction) algorithm \cite{rose2010automatic} at this time, and MatScholar, which uses name entity recognition (NER) along with an RNN-LSTM (recurrent neural network-long short term memory) structure. This comparison extended an exploratory, first-phase evaluation comparing these two applications that was reported at the the ACM/IEEE Joint Conference on Digital Libraries\cite{zhao2020scholarly}. For the 2021 study, material science researchers assessed the relevance of the results, and were asked for feedback on the two applications. For the HIVE4MAT application they indicated the ease in selecting relevant terminology for representing the content of scientific research articles or data sets they aimed publish. The combined relevant and partially relevant performance was 66\% as reported in Table \ref{tab:tel-paper} (next page).

\begin{table}[h!]
  \caption{Relevance Study Results for Inorganic Materials Literature}
  \label{tab:tel-paper}
  \begin{tabular}{lr}
    \toprule
    Evaluation Result of HIVE-4-MAT & \\
    \midrule
    Sample size (number of abstracts)            & 60           \\
    Number of extracted terms                    & 987          \\
    Relevant terms                               & 392          \\
    Partially Relevant terms                     & 261          \\
    Not Relevant terms                           & 334          \\
    Average terms extracted per abstract (range) & 16.45 (5-30) \\
    Percentage relevancy                         & 66.16\%   \\
    \bottomrule
  \end{tabular}
\end{table}

Researchers found HIVE4MAT was helpful in that it provided access to knowledge recorded in articles, as well as insight into a terminological structure. Another noted benefit was recognizing that HIVE4MAT's supports the reuse of existing ontologies and other semantic systems. This permits greater data interoperability and breaks down existing semantic silos. On the development end, HIVE4MAT does not require a large dataset and LLM training. In terms of HIVE4MAT's limitations, the knowledge extraction and indexing capabilities rely on the composition of terms in the ontology and the algorithm used, and the results lack the specificity found with MatScholar. 

The second evaluation tested the performance of HIVE4MAT's automatic indexing feature and focused Metal Organic Frameworks (MOFs) research. A sample of 10 articles reporting on MOF research were run through the HIVE4MAT, and processed against ten material science ontologies. A total of 282 candidate terms were returned, resulting in an average of 28 terms per article, and 28 terms per ontology. The articles had a standard deviation of 12 terms (Max 42, Min 3), and the vocabularies had a standard deviation of 23 terms (Max 67, Min 0). Five MOFs researchers participated in an evaluation and rated whether each of the 282 terms was relevant to it's corresponding article. The evaluators worked with a three-tiered evaluation system consisting of the following measures: relevant, partially relevant, or not relevant, and the result are reported in Table~\ref{tab:mofs-docs} (above) and Table~\ref{tab:mofs-onts} (next page). 

\begin{table}
  \caption{MOFs Relevance Study Results for Documents}
  \label{tab:mofs-docs}
  \begin{tabular}{lrrr}
    \toprule
    Article ID & Num Candidate Terms & Num Relevant Terms & Precision\\
    \midrule
    M1 & 17 & 8 & 47.06\% \\
    M2 & 42 & 21 & 50.00\% \\
    M3 & 26 & 15 & 57.69\% \\
    M4 & 3 & 0 & 0.00\% \\
    M5 & 36 & 11 & 30.56\% \\
    M6 & 41 & 11 & 26.83\% \\
    M7 & 27 & 10 & 37.04\% \\
    M8 & 23 & 8 & 34.78\% \\
    M9 & 32 & 12 & 37.50\% \\
    M10 & 35 & 14 & 40.00\% \\
    \bottomrule
  \end{tabular}
\end{table}

\begin{table}
  \caption{MOFs Relevance Study Results for Ontologies}
  \label{tab:mofs-onts}
  \begin{tabular}{lrrr}
    \toprule
    Ontology ID & Num Candidate Terms & Num Relevant Terms & Relevance\\
    \midrule
    AMONTOLOGY & 3 & 0 & 0.00\% \\
    BAO & 62 & 11 & 17.74\% \\
    BWMD-MID & 0 & 0 & 0.00\% \\
    CHMO & 28 & 7 & 25.00\% \\
    EMMO & 31 & 13 & 41.94\% \\
    MATONTO & 39 & 23 & 58.97\% \\
    MM & 67 & 35 & 52.24\% \\
    NMRRVOCAB & 25 & 12 & 48.00\% \\
    PROCCHEMICAL & 5 & 1 & 20.00\% \\
    USGS & 22 & 8 & 36.36\% \\
    \bottomrule
  \end{tabular}
\end{table}

Any candidate term that was deemed relevant or partially relevant by four or five of the five experts was considered relevant, while all other terms were considered not relevant as they did not meet this threshold. Of the 282 candidate terms corresponding to an article, 110 were considered relevant, giving a precision of 39\%. This is a vast improvement given that the relevancy results for some initial testing of HIVE4MAT's automatic indexing feature on MOF-related articles had performed extremely poorly. The initial improvements found during this evaluation were the result HIVE4MAT algorithm modification reported in \cite{polson_aligning_2024}.

Work is now underway to further streamline the process of automatically downloading ontologies from common repositories and adding updates to HIVE4MAT. This enhancement will both expedite and simplify HIVE4MAT updates and insure the most current versions of ontologies are employed for automatic indexing,  and accessible to users for searching search and navigation. This enhancement is a top priorities that is part of the HIVE4MAT development road map.

\section{Implications and Future Plans}

Preliminary testing of HIVE4MAT has been promising in selecting relevant terms for materials science \cite{zhao2021exploratory}. While some may raise an questions with translating an ontology into SKOS, there are benefits and other considerations that need to be considered. First, HIVE4MAT does not strive to be a registry and the original OWL ontology is still available for each ontology's respective host, thus maintaining its complexity.

Second, HIVE4MAT focuses on the human discovery by bringing together ontologies that have emerged as standards in the field of material science. The benefit of HIVE4MAT is allowing users to comfortably interact and explore rich collections of semantics relevant for indexing textual resources or textual components of resources that may have other associated media. Indeed, inference based features can be explored by other applications.

Third, OWL implementations vary across schemas, presenting significant mapping challenges. While SKOS may reduce relationship specificity, it allows for greater interoperability when investigating multiple ontologies at a first pass or when ontologies are being explored for simply indexing. Many ontologies utilize the RDF schema \verb|comment| as a field to annotate ontology concepts, but other ontologies within HIVE4MAT only use SKOS \verb|scopeNote|, or the Common Core Ontology \verb|definition|. While each ontology may use different fields for recording notes about a concept, translating these schemas all to SKOS within HIVE4MAT allows for schema to be interoperable at a general level, without changing the basic hierarchy of the original ontology.

Once research and implementation on automatic ontology ingestion into HIVE4MAT is complete, the HIVE4MAT team will turn their attention to other development roadmap plans. First, the front end will be overhauled so that the back end logic and the front end logic can be easier to maintain and further develop. This will make it easier to add new features, and for users to access the code and implement similar features in their own systems. Second, we  will implement a new feature which will allow an "ontology to ontology" search. The goal is to allow a user to input a list of terms they use at an individual, project, institution, or discipline level, and see if that list of terms has a best match with an existing ontology, or a set of ontologies. While there is a growing body of work on ontology alignment\cite{xue2023generative,he2022bertmap,lateef2024optimized}, applications supporting easy "ontology to ontology" search appears limited. Third, we aim to provide more provenance metadata about each ontology in the HIVE4MAT system. Each ontology has a context for when and where the ontology was created, who (individual or group) created the ontology, and where it's used. Making that historical context more visible to users may help with discerning which ontologies a project should consider implementing.

Overall our development roadmap will contribute to more robust HIVE4MAT infrastructure. As HIVE4MAT advances we will more widely share our developments with the materials science community, and other communities seeking to leverage multiple ontologies.

\begin{acknowledgments}
We acknowledge the support of NSF-OAC\#2118201  
\end{acknowledgments}


\section{Online Resources}
\begin{itemize}
\item
  \href{http://bioportal.bioontology.org/}{NCBO BioPortal}
\item 
  \href{http://www.obofoundry.org/}{OBO Foundry}
\item 
  \href{https://materials.registry.nist.gov/}{NIST Materials Registry}
\item 
  \href{https://www.industrialontologies.org/}{Industrial Ontology Foundry}
\item 
  \href{https://matportal.org/ontologies}{MatPortal}
\item 
  \href{https://www.mines.edu/id4/}{NSF/HDR-ID4}
\end{itemize}


\begin{thebibliography}{10}
\expandafter\ifx\csname url\endcsname\relax
  \def\url#1{\texttt{#1}}\fi
\expandafter\ifx\csname urlprefix\endcsname\relax\def\urlprefix{URL }\fi
\expandafter\ifx\csname href\endcsname\relax
  \def\href#1#2{#2} \def\path#1{#1}\fi

\bibitem{bayerlein2024pmd}
B.~Bayerlein, M.~Schilling, H.~Birkholz, M.~Jung, J.~Waitelonis, L.~M{\"a}dler,
  H.~Sack, Pmd core ontology: Achieving semantic interoperability in materials
  science, Materials \& Design 237 (2024) 112603.

\bibitem{zhao2018nanomine}
H.~Zhao, Y.~Wang, A.~Lin, B.~Hu, R.~Yan, J.~McCusker, W.~Chen, D.~L.
  McGuinness, L.~Schadler, L.~C. Brinson, Nanomine schema: An extensible data
  representation for polymer nanocomposites, APL Materials 6~(11).

\bibitem{wilkinson2016fair}
M.~D. Wilkinson, M.~Dumontier, I.~J. Aalbersberg, G.~Appleton, M.~Axton,
  A.~Baak, N.~Blomberg, J.-W. Boiten, L.~B. da~Silva~Santos, P.~E. Bourne,
  et~al., The fair guiding principles for scientific data management and
  stewardship, Scientific data 3~(1) (2016) 1--9.

\bibitem{berners2001publishing}
T.~Berners-Lee, J.~Hendler, Publishing on the semantic web, Nature 410~(6832)
  (2001) 1023--1024.

\bibitem{bizer2009emerging}
C.~Bizer, The emerging web of linked data, IEEE intelligent systems 24~(5)
  (2009) 87--92.

\bibitem{Miles:09:SSK}
A.~Miles, S.~Bechhofer, {SKOS} simple knowledge organization system reference,
  {W3C} recommendation, W3C,
  https://www.w3.org/TR/2009/REC-skos-reference-20090818/ (Aug. 2009).

\bibitem{DimaAlden2016IIft}
A.~Dima, S.~Bhaskarla, C.~Becker, M.~Brady, C.~Campbell, P.~Dessauw,
  R.~Hanisch, U.~Kattner, K.~Kroenlein, M.~Newrock, A.~Peskin, R.~Plante, S.-Y.
  Li, P.-F. Rigodiat, G.~S. Amaral, Z.~Trautt, X.~Schmitt, J.~Warren,
  S.~Youssef, Informatics infrastructure for the materials genome initiative,
  JOM (1989) 68~(8) (2016) 2053--2064.

\bibitem{alma991010609049704721}
Materials Genome Initiative strategic plan, Executive Office of the President,
  National Science and Technology Council, Committee on Technology,
  Subcommittee on the Materials Genome Initiative, Washington, District of
  Columbia, 2014.

\bibitem{GoldbeckG2019Arla}
G.~Goldbeck, E.~Ghedini, A.~Hashibon, G.~Schmitz, J.~Friis, A reference
  language and ontology for materials modelling and interoperability, 2019.

\bibitem{VoigtSvenP.2021Mgo}
S.~P. Voigt, S.~R. Kalidindi, Materials graph ontology, Materials letters 295
  (2021) 129836--.

\bibitem{DavidMrdjenovich2020pAKG}
D.~Mrdjenovich, M.~K. Horton, J.~H. Montoya, C.~M. Legaspi, S.~Dwaraknath,
  V.~Tshitoyan, A.~Jain, K.~A. Persson, propnet: A knowledge graph for
  materials science, Matter 2~(2) (2020) 464--480.

\bibitem{MaBoran2023MUoP}
B.~Ma, N.~J. Finan, D.~Jany, M.~E. Deagen, L.~S. Schadler, L.~C. Brinson,
  Machine-learning-assisted understanding of polymer nanocomposites
  composition–property relationship: A case study of nanomine database,
  Macromolecules 56~(11) (2023) 3945--3953.

\bibitem{whetzel2011bioportal}
P.~L. Whetzel, N.~F. Noy, N.~H. Shah, P.~R. Alexander, C.~Nyulas, T.~Tudorache,
  M.~A. Musen, Bioportal: enhanced functionality via new web services from the
  national center for biomedical ontology to access and use ontologies in
  software applications, Nucleic acids research 39~(suppl\_2) (2011)
  W541--W545.

\bibitem{smith2007obo}
B.~Smith, M.~Ashburner, C.~Rosse, J.~Bard, W.~Bug, W.~Ceusters, L.~J. Goldberg,
  K.~Eilbeck, A.~Ireland, C.~J. Mungall, et~al., The obo foundry: coordinated
  evolution of ontologies to support biomedical data integration, Nature
  biotechnology 25~(11) (2007) 1251--1255.

\bibitem{draxl2018nomad}
C.~Draxl, M.~Scheffler, Nomad: The fair concept for big data-driven materials
  science, Mrs Bulletin 43~(9) (2018) 676--682.

\bibitem{PlanteRaymondL2021IaRF}
R.~L. Plante, C.~A. Becker, A.~Medina-Smith, K.~Brady, A.~Dima, B.~Long, L.~M.
  Bartolo, J.~A. Warren, R.~J. Hanisch, Implementing a registry federation for
  materials science data discovery, Data science journal 20~(1).

\bibitem{jonquet2023ontoportal}
C.~Jonquet, S.~Bouazzouni, J.~Graybeal, J.~L. Vendetti, Ontoportal workshop
  2023 report, Ph.D. thesis, INRAE, Universit{\'e} de Montpellier, Stanford
  University (2023).

\bibitem{DraxlClaudia2019TNlf}
C.~Draxl, M.~Scheffler, The nomad laboratory: from data sharing to artificial
  intelligence, JPhys materials 2~(3) (2019) 36001--.

\bibitem{national_information_standards_organization_guidelines_2010}
N.~I.~S. Organization, Guidelines for the construction, format, and management
  of monolingual controlled vocabulary, {OCLC}: 59757037.
\newblock \href {http://dx.doi.org/10.3789/ansi.niso.z39.19-2005R2010}
  {\path{doi:10.3789/ansi.niso.z39.19-2005R2010}}.

\bibitem{baker2013key}
T.~Baker, S.~Bechhofer, A.~Isaac, A.~Miles, G.~Schreiber, E.~Summers, Key
  choices in the design of simple knowledge organization system (skos), Journal
  of Web Semantics 20 (2013) 35--49.

\bibitem{rose2010automatic}
S.~Rose, D.~Engel, N.~Cramer, W.~Cowley, Automatic keyword extraction from
  individual documents, Text mining: applications and theory (2010) 1--20.

\bibitem{campos2020yake}
R.~Campos, V.~Mangaravite, A.~Pasquali, A.~Jorge, C.~Nunes, A.~Jatowt, Yake!
  keyword extraction from single documents using multiple local features,
  Information Sciences 509 (2020) 257--289.

\bibitem{zhao2021exploratory}
X.~Zhao, J.~Greenberg, V.~Meschke, E.~Toberer, X.~Hu, An exploratory analysis:
  extracting materials science knowledge from unstructured scholarly data, The
  Electronic Library 39~(3) (2021) 469--485.

\bibitem{zhao2020scholarly}
X.~Zhao, Scholarly big data: computational approaches to semantic labeling in
  materials science, in: IEEEACM Joint Conference on Digital Libraries JCDL,
  2020.

\bibitem{polson_aligning_2024}
K.~Polson, J.~Greenberg, S.~McClellan, \href{https://osf.io/2kub9/}{Aligning
  {Keywords} from {Long} {Form} {Prose} to {Controlled} {Vocabulary}},
  publisher: OSF (May 2024).
\newline\urlprefix\url{https://osf.io/2kub9/}

\bibitem{xue2023generative}
X.~Xue, Q.~Huang, Generative adversarial learning for optimizing ontology
  alignment, Expert Systems 40~(4) (2023) e12936.

\bibitem{he2022bertmap}
Y.~He, J.~Chen, D.~Antonyrajah, I.~Horrocks, Bertmap: a bert-based ontology
  alignment system, in: Proceedings of the AAAI Conference on Artificial
  Intelligence, Vol.~36, 2022, pp. 5684--5691.

\bibitem{lateef2024optimized}
A.~Lateef Haroon~PS, S.~N. Patil, P.~Bidare~Divakarachari, P.~Falkowski-Gilski,
  M.~Rafeeq, An optimized system for sensor ontology meta-matching using swarm
  intelligent algorithm, Internet Technology Letters (2024) e498.

\end{thebibliography}
\end{document}